\documentclass[final,5p,times,twocolumn]{elsarticle}

\usepackage{amsmath,amssymb,amsfonts}
\usepackage{mathtools}
\usepackage{lineno,hyperref}

\begin{document}

\begin{frontmatter}

\title{Ultra low noise readout with travelling wave parametric amplifiers: the DARTWARS project}

\author[lnf]{A. Rettaroli\corref{mycorrespondingauthor}}
\cortext[mycorrespondingauthor]{Corresponding author. Email: alessio.rettaroli@lnf.infn.it}

\author[unisa,infn-sa]{C. Barone}
\author[unimib,infn-mi]{M. Borghesi}
\author[unimib,infn-mi]{S. Capelli}
\author[unisa,infn-sa]{G. Carapella}
\author[unisalento,infn-le]{A. P. Caricato}
\author[ino-cnr,unitn]{I. Carusotto}
\author[fbk,infn-tifpa]{A. Cian}
\author[lnf]{D. Di Gioacchino}
\author[inrim,infn-tifpa]{E. Enrico}
\author[fbk,infn-tifpa,ifn-cnr]{P. Falferi}
\author[inrim,polito]{L. Fasolo}
\author[unimib,infn-mi]{M. Faverzani}
\author[infn-mi]{E. Ferri}
\author[infn-sa,unisannio-sci]{G. Filatrella}
\author[lnf]{C. Gatti}
\author[unimib,infn-mi]{A. Giachero}
\author[fbk,infn-tifpa]{D. Giubertoni}
\author[unisa,infn-sa]{V. Granata}
\author[inrim,polito]{A. Greco\fnref{myfootnote}}
\author[unisa,infn-sa]{C. Guarcello}
\author[unimib,infn-mi]{D. Labranca}
\author[unisalento,infn-le]{A. Leo}
\author[lnf]{C. Ligi}
\author[lnf]{G. Maccarrone}
\author[fbk,infn-tifpa]{F. Mantegazzini}
\author[fbk,infn-tifpa]{B. Margesin}
\author[unisalento,infn-le]{G. Maruccio}
\author[infn-sa]{C. Mauro}
\author[unitn,infn-tifpa]{R. Mezzena}
\author[unisalento,infn-le]{A. G. Monteduro}
\author[unimib,infn-mi]{A. Nucciotti}
\author[inrim,infn-tifpa]{L. Oberto}
\author[unimib,infn-mi]{L. Origo}
\author[unisa,infn-sa]{S. Pagano}
\author[infn-sa,unisannio-eng]{V. Pierro}
\author[lnf]{L. Piersanti}
\author[inrim,infn-to]{M. Rajteri}
\author[unisalento,infn-le]{S. Rizzato}
\author[fbk,infn-tifpa,ifn-cnr]{A. Vinante}
\author[unimib,infn-mi]{M. Zannoni}

\address[lnf]{INFN - Laboratori Nazionali di Frascati, Via Enrico Fermi, I-00044, Frascati, Italy}
\address[infn-le]{INFN – Sezione di Lecce, Via per Arnesano, I-73100 Lecce, Italy}
\address[unisalento]{University of Salento, Department of Physics, Via per Arnesano, I-73100 Lecce, Italy}
\address[unimib]{University of Milano Bicocca, Department of Physics, Piazza della Scienza , I-20126 Milano, Italy}
\address[infn-mi]{INFN - Milano Bicocca, Piazza della Scienza, I-20126 Milano, Italy}
\address[unisa]{University of Salerno, Department of Physics, Via Giovanni Paolo II, I-84084 Fisciano, Salerno, Italy}
\address[infn-sa]{INFN - Napoli, Salerno group, Via Giovanni Paolo II, I-84084 Fisciano, Salerno, Italy}
\address[unisannio-sci]{University of Sannio, Department of Science and Technology, via Francesco de Sanctis, I-82100, Benevento, Italy}
\address[unisannio-eng]{University of Sannio, Department of Engineering, Corso Garibaldi, I-82100 Benevento, Italy}
\address[ino-cnr]{INO-CNR BEC Center,  Via Sommarive, I-38123 Povo, Italy}
\address[unitn]{University of Trento, Department of Physics, Via Sommarive, I-38123, Povo, Trento, Italy}
\address[fbk]{Fondazione Bruno Kessler, Via Sommarive, I-38123, Povo, Trento, Italy}
\address[infn-tifpa]{INFN - TIFPA, Via Sommarive, I-38123, Povo, Trento, Italy}
\address[ifn-cnr]{IFN-CNR , Via Sommarive, I-38123 Povo, Trento, Italy}
\address[inrim]{INRiM - Istituto Nazionale di Ricerca Metrologica, Strada delle Cacce, I-10135 Turin, Italy}
\address[polito]{Politecnico di Torino, Corso Duca degli Abruzzi, I-10129 Turin, Italy}
\address[infn-to]{INFN - Torino, Via Pietro Giuria, I-10125 Turin, Italy}

\fntext[myfootnote]{Currently at: National Enterprise for nanoscience and Technology –- NEST, Istituto Nanoscienze -- CNR and Scuola Normale Superiore, Pisa, Italy}

\begin{abstract}
The DARTWARS project has the goal of developing high-performing innovative travelling wave parametric amplifiers with high gain, large bandwidth, high saturation power, and nearly quantum-limited noise. The target frequency region for its applications is $5-10$~GHz, with an expected noise temperature of about 600~mK. The development follows two different approaches, one based on Josephson junctions and one based on kinetic inductance of superconductors. This contribution mainly focuses on the Josephson travelling wave parametric amplifier, presenting its design, preliminary measurements and the test of homogeneity of arrays of Josephson junctions.
\end{abstract}


\end{frontmatter}

\section{Introduction}

Ultra-low noise detection near the quantum limit and amplification over a large bandwidth are fundamental requirements in forthcoming particle physics applications operating at low temperatures, such as neutrino measurements, x-ray observations, CMB measurements, and light dark matter detection, as well as in quantum computing applications, where the high-fidelity readout is key. In these fields, arrays of detectors are being used, such as arrays of MKIDs (microwave kinetic inductance detectors), arrays of TESs (transition edge sensors), microwave resonant cavities and arrays of qubits, all requiring multiplexed readout.

The readout sensitivity of these detectors is currently limited by the noise temperature and bandwidth of available cryogenic amplifiers such as HEMTs (high-electron-mobility transistors) or JPAs (Josephson parametric amplifiers). Comparing the two technologies, HEMTs have the advantages of providing high gain ($>30$~dB), large bandwidth (few  GHz) and high dynamic range, while JPAs have typical gain of about 20~dB, a small dynamic range ($<-100$~dBm) and a small instantaneous bandwidth ($\sim 100$~MHz). Nevertheless, JPAs allow to significantly boost the sensitivity of the detection since their added noise reaches the quantum limit, whereas HEMTs noise is 10--40 times above that limit.

DARTWARS (Detector Array Readout with Travelling Wave AmplifieRS) aims to develop a device with large bandwidth and nearly quantum-limited noise at the same time, exploiting the concept of parametric amplification with microwaves travelling along a transmission line with embedded superconducting nonlinear lumped elements.

\section{The project}
A parametric amplifier is a type of parametric oscillator, which is a harmonic oscillator where its parameters are varied with time:
\begin{equation}
\frac{\text{d}^2x}{\text{d}t^2} + \beta(t) \frac{\text{d}x}{\text{d}t} + \omega^2(t) \, x = 0 .
\end{equation}
$\beta(t)$ and $\omega(t)$ are the damping coefficient and the resonance frequency. If these parameters are varied at about twice the resonance frequency by a pump signal, the oscillator absorbs energy from it.
If the loss is not sufficient to dampen this energy, the oscillations grow exponentially, whereas, below this limit, the pumped energy is transferred to the signal that is amplified.

A TWPA is designed as a transmission line with tunable embedded reactive elements, as inductances. The nonlinear inductance is implemented by means either of Josephson junctions (JJs) or the kinetic inductance (KI) of superconductors, for which the relationship with the current is, at the first order, $L(I) \simeq L_0 (1 + (I/I_c)^2)$, where in JJs $I_c$ is the junctions' critical current, while in KI devices $I_c$ is the superconductor critical current. Both three-wave mixing (3WM) and four-wave mixing (4WM) are possible: in 3WM a single pump photon converts into signal and idler photons, $\omega_p = \omega_s+\omega_i$, whereas in 4WM two pump photons convert into signal and idler photons, $2\omega_p = \omega_s+\omega_i$ \cite{malnouPRX}.

The development of Josephson travelling wave parametric amplifiers (JTWPAs) and of kinetic inductance TWPAs (KITWPAs) has been already investigated and demonstrated, as for example in \cite{oBrien,jtwpa2} and \cite{eom,kitwpa2} respectively. In these cases, gain values up to 15~dB over a bandwidth of a few GHz have been reached, but significant progress is still needed. The purpose of DARTWARS is to investigate the fabrication of both JTWPA and KITWPA devices \cite{giachero}.

\begin{figure}
\centering
\includegraphics[width=0.36\textwidth]{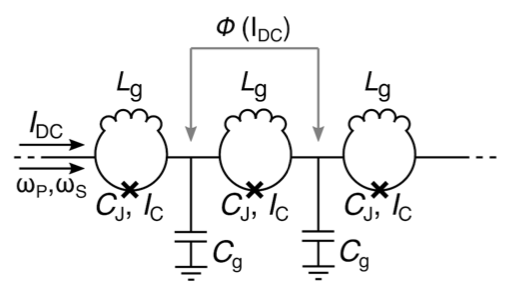}
\caption{Circuit schematic of the Josephson metamaterial. The central conductor of the coplanar waveguide is made by a chain of rf-SQUIDs composed by a geometrical inductor $L_g$ and a Josephson junction of critical current $I_c$ and capacitance $C_J$. The line is capacitively shunted to ground by capacitors $C_g$.}
\label{fig:designJTWPA}
\end{figure}

JTWPAs are made of a coplanar waveguide (CPW) embedding a chain of rf-SQUIDs (see Fig.~\ref{fig:designJTWPA}), which can be biased by a DC current or a magnetic field to activate the 3WM or 4WM nonlinearities. The design follows the coupled mode equation approach developed by INRiM \cite{grecoPRB,fasoloTASC}. To avoid power leakage into higher frequency tones, the CPWs are equipped with a modified dispersion relation following two different approaches, the Resonant Phase Matching (RPM) and the Quasi-Phase Matching (QPM). RPM uses a reduced plasma frequency mixed to a periodic load of LC resonators to create mismatch among the travelling tones, in order to suppress higher harmonic generation, then the signal tone that is meant to be amplified is re-phased through the opening of a bandgap in the dispersion relation. On the contrary, QPM uses a mix of low plasma frequency and a sign modulation of the nonlinearity into the medium to suppress higher harmonic generation and stimulate amplification by changing the phase of the travelling waves of $\pi$ after a coherence length has been reached.

\begin{figure}
\centering
\includegraphics[width=0.32\textwidth]{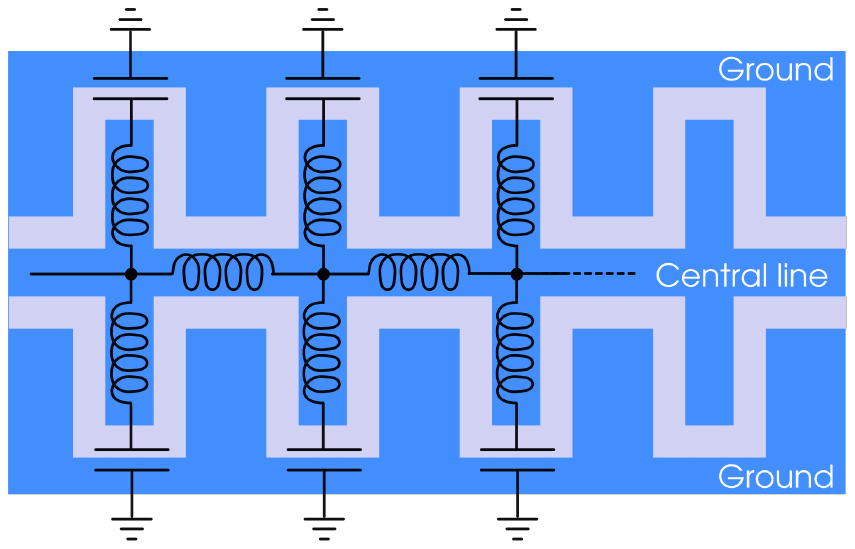}
\caption{Circuital schematic of a unit cell of a superconducting artificial line made of CPW sections. Each cell is composed by a series inductance $L_d$ and two resonators with inductance $L_f$ and capacitance to ground $C/2$.}
\label{fig:designKITWPA}
\end{figure}

In the KITWPA devices (Fig.~\ref{fig:designKITWPA}), the momentum conservation can be achieved in two ways: \emph{a)} dispersion engineering the CPW with periodic loadings creating a frequency gap; \emph{b)} building an artificial transmission line, that uses lumped-element inductors and capacitors. The characteristic impedance of the transmission line is modified every one-sixth of a wavelength at a frequency slightly above the pump frequency $f_p$ to form a wide stopband at $3f_p$, allowing suppression of the third armonic of the pump. In addition, every third loading is modified in length (longer or shorter relative to the first two) to create a narrow stopband near $f_p$; this allows the pump to fulfill the phase-matching condition. The CPW-type amplifiers are made of NbTiN and are meter-long transmission lines, causing impedance mismatches which are the likely cause of large ripples in the gain profile, with the result that the amplifiers suffer from self-heating due to the strong pump tone, creating an excess of thermal noise. On the other hand, the lumped-element approach brings the advantage of having a shorter transmission line resulting in a higher fabrication yield.

The goals of the DARTWARS project within 2024 are: \emph{a)} the development of high-performing parametric amplifiers by exploring new design solutions, new materials and fabrication processes, to achieve a gain value around 20~dB, a high saturation power (around $-50$~dBm), a large bandwidth (in the 5--10~GHz range), noise near the quantum limit ($T_n \lesssim 600$~mK at these frequencies) and to reduce the gain ripples; \emph{b)} the readout demonstration of different detectors and devices involved in the next-generation particle physics experiments, such as MKIDs, TESs, microwave resonant cavities and qubits.

\section{Preliminary measurements on JTWPA protorype}
\label{sec:measurements}
The device was fabricated by INRiM and tested at LNF in a dry dilution refrigerator with the lowest temperature stage at $T=15$~mK. It is composed of 15 sections of coplanar waveguide embedding 990 nonhysteretic rf-SQUIDs connected by bended sections of CPW. The values of the circuit parameters of the Josephson metamaterial, by design, are a ground capacitance $C_g= 13.0$~fF, a geometrical inductance $L_g= 45$~pH, a Josephson capacitance $C_J= 25.8$~fF and a  Josephson critical current $I_c= 1.5~\mu$A. The Josephson junctions were fabricated exploiting an electron beam lithography process on a  double layer polymeric mask, followed by an Aluminum e-gun evaporation.

The characterization of the JTWPA consists in evaluating its 3WM capabilities and its gain through pump-on pump-off measurements. Two-tones measurements are possible by supplying in input a weak signal tone and a driving pump tone, coupled together by a directional coupler. The stage of amplification is composed by a low-noise cryogenic HEMT (put at 4~K) and a room-temperature FET, providing a gain of 30~dB each. An rf splitter allows to send the output both to a spectrum analyzer, to perform power spectra, and to a vector network analyzer, to measure scattering parameters. Finally, a current generator connected to the device through bias tees provides the DC current bias to the device.
(More details in \cite{jtwpaMisure}).

\begin{figure}
\centering
\includegraphics[width=0.42\textwidth]{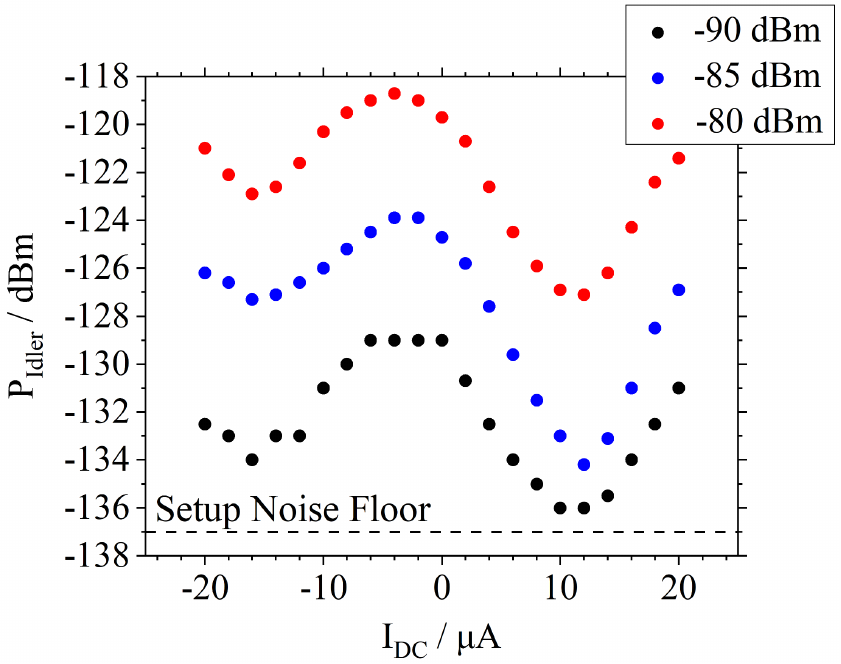}
\caption{Modulation of the output idler tone power ($P_{\text{Idler}}$) generated via 3WM as a function of the bias current $I_{\text{DC}}$. Here the JTWPA is excited with a signal tone at frequency $\nu_s = 3.3$~GHz , for three different values of the driving pump tone at frequency $\nu_p = 6.75$~GHz.}
\label{fig:3wm}
\end{figure}

Nonlinear effects generate idler tones that have different frequencies depending on the order of nonlinearity that causes them. Fig.~\ref{fig:3wm} reports the power of the output idler tone $P_{\text{Idler}}$, generated via 3WM, as a function of the DC bias current $I_{\text{DC}}$. The pump tone is at $\nu_p = 6.75$~GHz, with three different power values ($-90$, $-85$ and $-80$~dBm), and the signal tone is at $\nu_s = 3.3$~GHz with a power of $P_s=-67$~dBm. For this mixing process the idler is generated at $\nu_i = \nu_p - \nu_s = 3.45$~GHz. The 3WM idler should present a minimum at zero $I_{\text{DC}}$, as expected from the Kerr nonlinearity of an rf-SQUID, but here we note a shift of the minima, which is attributed to magnetic field trapping during the cooling of the dilution refrigerator.  Moreover, the suppression of the 3WM idler tone is not complete, since the data in Fig.~\ref{fig:3wm} do not reach the noise floor of the setup (dashed line). It has to be noticed that the modulation of the 3WM process here reported is limited to around 10 dB, since it is reasonably affected by nonidealities of the JTWPA and the surrounding environment.

\begin{figure}[h]
\centering
\includegraphics[width=0.45\textwidth]{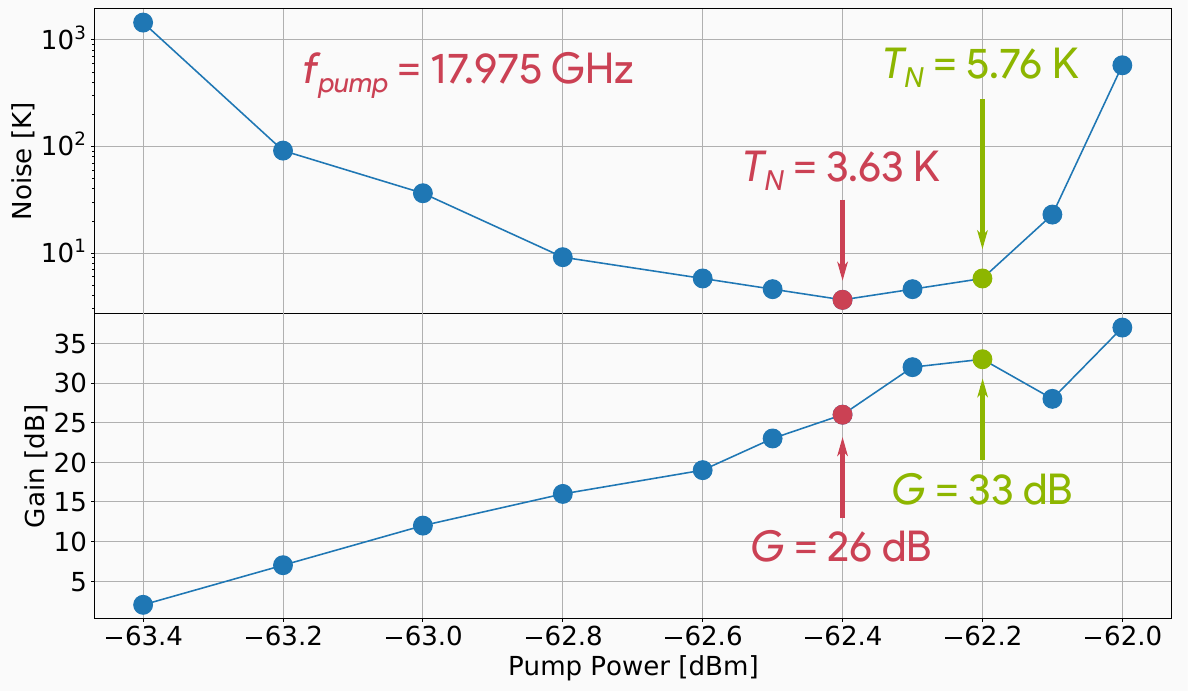}
\caption{\emph{Top:} Noise temperature as a function of the input pump power for frequencies $\nu_p=17.975$~GHz and $\nu_s=9$~GHz (nondegenerate mode). \emph{Bottom:} Signal gain as a function of the pump power with the same tone frequencies, evaluated with the pump-on pump-off technique.}
\label{fig:gain-noise}
\end{figure}

Then, parametric amplification has been quantified with gain measurements, by means of the pump-on pump-off technique. The gain is studied both in the degenerate ($\nu_p=\nu_s$) and nondegenerate mode, as a function of the pump power. Although we do not observe a constant gain over a large bandwidth, values between 25 and 30~dB are reached for particular values of pump frequency and power, as shown in Fig.~\ref{fig:gain-noise} for $\nu_p=17.975$~GHz and $\nu_s=9$~GHz (nondegenerate mode). Unfortunately, the actual noise temperature of the JTWPA could not be estimated due to a malfunctioning of attenuators at very low temperatures, causing an excess thermal noise (the minimum measured noise temperature was $T_n=3.63$~K).

After the prototype was characterized, a study of the homogeneity of the Josephson junctions fabricated with the same process was carried out, to improve the performances of future devices. For this reason, a sample of 960 JJs with critical current $I_c=4~\mu$A, self-capacitance $C=225$~fF and expected normal resistance $R_n\simeq 80~\Omega$ was fabricated. Two oxidation techniques were used, a dynamic oxidation with an $O_2$ pressure of $4.30\times 10^{-4}$~Torr for a time of 660~s, and a static oxidation at $1.58\times 10^{-3}$~Torr for 344~s. The two processes should bring to similar oxide barrier thicknesses and similar resistances.

\begin{figure}
\centering
\includegraphics[width=0.35\textwidth]{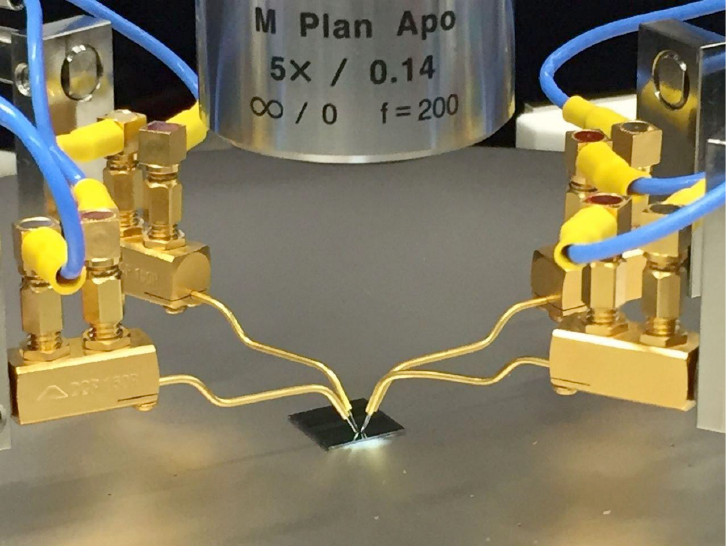}
\caption{Probe station used to perform four-terminal resistance measurements on the sample of 960 Josephson junctions.}
\label{fig:probe}
\end{figure}

The normal resistances of JJs were tested with a probe station, which performs four-terminal measurements (see Fig.~\ref{fig:probe}). As a result, the resistance values are distributed around about $12~\Omega$, with a spread around 5\%$-$10\%. There are also ascending or descending gradients depending on what is the position of the arrays of junctions along the wafer. Finally, junctions fabricated with the static oxidation process show higher overall resistances than the dynamic oxidation ones.

\section{Conclusions}
TWPAs are promising candidates of quantum-limited microwave amplifiers for applications in fundamental physics and quantum computing. DARTWARS aims at developing (nearly-)quantum limited TWPAs exploiting Josephson junctions and kinetic inductance of superconductors, exploring new designs and materials, and demonstrating the readout of several devices (TES/MKIDs/RF cavities/qubits).
With the preliminary characterization of a prototype of JTWPA we demonstrated the 3WM modulation, although with some nonhomogeneities, as well as good capabilities of parametric amplification, measuring gain values between 25 and 30~dB for particular frequencies.

The results show that there is room for improvement. In fact, INRiM is committed to implement the design with the modified dispersion relation given by the RPM technique in the JTWPA structure, to reduce the phase mismatch between the travelling tones. Furthermore, numerical studies on the QPM approach are being performed. On the fabrication side, a new realization technique is being investigated, consisting in a two-step lithography; moreover, to reduce the single JJ areas nonhomogeneity due to the overlap of unpredictable rounded edges, a new design exploiting a double-layer mask is under development. New tests of the homogeneity of samples of JJs are being performed, without observing better results with respect to Section~\ref{sec:measurements} at the moment.

Finally, progress has been made in the development of KITWPAs. In fact, NbTiN patterned into micro-resonators was characterized to estimate the kinetic inductance of the material and its nonlinearity. The kinetic inductance was evaluated measuring the resonance frequency of the resonators, and then it was related to the nonlinearity in the current. The KITWPA device design is close to completion, and the first prototype is foreseen for summer 2022.

\section*{Acknowledgments}
This work is supported by the Italian Institute of Nuclear Physics (INFN), within the Technological and Interdisciplinary research commission (CSN5), by the European Union's H2020-MSCA Grant Agreement No. 101027746, and by the Joint Research Project PARAWAVE of the European Metrology Programme for Innovation and Research (EMPIR). PARAWAVE received funding from the EMPIR programme co-financed by the Participating States and from the European Union's Horizon 2020 research and innovation programme.

\section*{References}

\bibliography{dartwars}

\end{document}